\documentclass[aps,12pt,reprint,groupedaddress,superscriptaddress,onecolumn,longbibliography,pra]{revtex4-2}
\usepackage[usenames,dvipsnames]{xcolor}
\usepackage{amssymb}
\usepackage{graphicx}
\usepackage{amsmath}
\usepackage{amsthm}
\usepackage{dsfont}
\usepackage[bookmarks=true,colorlinks,citecolor=blue,urlcolor=blue]{hyperref}
\usepackage{braket}
\usepackage{babel}
\usepackage{physics}
\usepackage{siunitx}
\usepackage{bm}
\usepackage{tikz}
\usetikzlibrary{arrows.meta,positioning,calc}
\graphicspath{{./}}

\begin{document}

\title{Fault-tolerant cost of shallow QAOA on near-symmetric optimization problems}

\author{Jernej Rudi Fin\v{z}gar}
\email{jernej-rudi.finzgar@iqm.tech}
\author{Martin Leib}
\author{Elisabeth Wybo}
\affiliation{IQM Quantum Computers, Georg-Brauchle-Ring 23-25, 80992 München, Germany}
\date{\today}

\begin{abstract}
Montanaro and Zhou~\cite{montanaroQuantumSpeedupsSolving2025} proved that depth-one Quantum Approximate Optimization Algorithm (QAOA) finds the planted solution of certain near-symmetric constraint satisfaction problems with probability $\Omega(1)$, and observed exponential runtimes for general-purpose classical solvers on explicit realizations, providing strong evidence of an empirical exponential speedup of low-depth QAOA. We ask what such a circuit costs fault-tolerantly. Although the cost Hamiltonian carries $\Theta(n^\ell)$ clauses, the rotation angle at which QAOA succeeds shrinks as $n^{1-\ell}$, and we show that small-angle Clifford$+T$ rotation synthesis reduces the non-Clifford cost per circuit to $\widetilde O(n^2)$ for every clause locality $\ell$. Compiling the circuit, however, requires the explicit clause list. We find that the mechanism underlying constant QAOA success fixes the degree-one Fourier coefficients of the cost, whose signs reveal the planted solution. However, this leakage need not reveal the solution: we design families in which the quadratic non-Clifford scaling and hard exact optimization coexist.
\end{abstract}

\maketitle

\section{Introduction}
\label{sec:introduction}

Quantum algorithms offer several routes to speedups in combinatorial optimization~\cite{abbasChallengesOpportunitiesQuantum2024}. Rigorous speedups are typically quadratic, as in minimum finding~\cite{durrQuantumAlgorithmFinding1999} and quantum branch-and-bound~\cite{montanaroQuantumSpeedupBranchandbound2020}, while superpolynomial separations have been demonstrated for certain structured optimization problems~\cite{pirnayInprincipleSuperpolynomialQuantum2024, jordanOptimizationDecodedQuantum2025}. For low-depth algorithms on explicit instances, evidence for larger speedups is so far predominantly empirical~\cite{shaydulinEvidenceScalingAdvantage2024,boulebnaneSolvingBooleanSatisfiability2024}. Montanaro and Zhou~\cite{montanaroQuantumSpeedupsSolving2025} provide a distinctive combination of these perspectives. They construct Constraint Satisfaction Problems (CSPs) with cost functions that are symmetric under permutations of variables relative to a planted solution, and show that for suitable cost function profiles sampling from depth-one Quantum Approximate Optimization Algorithm (QAOA) state finds the planted solution with probability $\Omega(1)$. In the value-oracle model, this yields a quantum-classical separation between $O(1)$ quantum queries and $\Omega(n/\log n)$ classical queries required to recover the solution. They further show that the QAOA guarantee survives random sparsification of the clause set, which breaks the exact symmetry. Interestingly, for the resulting near-symmetric instances, all classical solvers that they benchmark exhibit apparently exponential runtime scaling.

This empirical speedup with respect to strong classical solvers motivates the question: what resources are required to implement these constructions on early fault-tolerant quantum computers? At first sight the answer is discouraging. The relevant cost Hamiltonians contain $\Theta(n^\ell)$ clauses, with $\ell\ge 5$ in the hardest examples where a significant speedup was observed, and each clause contributes a phase-separator rotation requiring Clifford+$T$ synthesis. Counted naively, a single QAOA layer then demands $\widetilde O(n^\ell)$ non-Clifford gates, placing the relevant problem sizes beyond the reach of early fault-tolerant hardware.

However, the conditions for constant success probability of depth-one QAOA require the phase-separator angle to scale as $n^{1-\ell}$. Combining this with small-angle Clifford+$T$ synthesis~\cite{botheMoreEfficientClifford+T2026}, we find in Sec.~\ref{sec:cheap} that the non-Clifford cost per circuit is surprisingly only
$
    \widetilde O(n^2),
$
regardless of the clause locality $\ell$ and the sparsification rate, allowing considerably larger problem sizes to fit on early fault-tolerant devices (see Fig.~\ref{fig:ft-cost}).

A gate-level implementation, however, requires access to the explicit clause representation rather than only to an opaque oracle (see Fig.~\ref{fig:opening-oracle})~\cite{stoudenmireOpeningBlackBox2024, aaronsonReadFinePrint2015}. This additional access makes the low-degree Fourier coefficients of the cost directly computable from the clause list. For the planted constructions of Ref.~\cite{montanaroQuantumSpeedupsSolving2025}, we show in Sec.~\ref{sec:leak} that the same condition that makes the angle small fixes these degree-one coefficients so that their signs reveal the planted solution.

We show in Sec.~\ref{sec:escape} that small-angle QAOA can coexist with NP-hard exact optimization. We construct near-symmetric instances without a planted solution whose globally favorable sector is an exponentially large Hamming shell, while selecting an exact optimum within that shell remains NP-hard. The macroscopic landscape is deceptive, trapping local search away from the favorable sector. Nevertheless, depth-one QAOA at a small phase angle concentrates its output on this shell, while retaining the same \(\widetilde O(n^2)\) non-Clifford cost per circuit.

\definecolor{oracleblue}{RGB}{52,92,140}
\definecolor{quantumblue}{RGB}{45,105,170}
\definecolor{classicalred}{RGB}{172,62,62}
\definecolor{softgray}{RGB}{245,246,248}
\definecolor{linegray}{RGB}{105,110,120}
\definecolor{green}{RGB}{62,140,92}

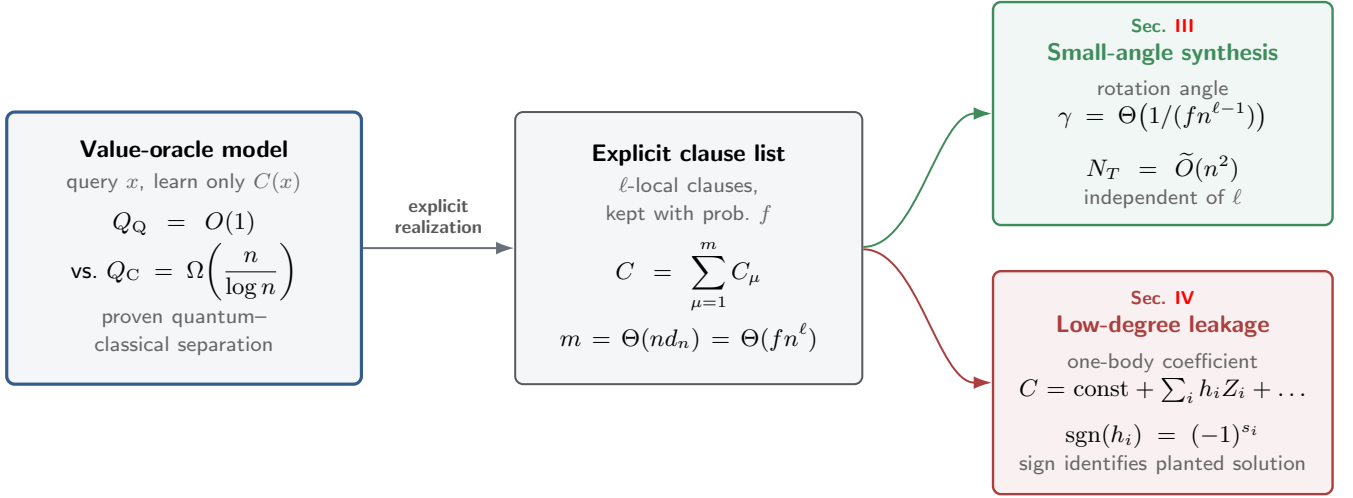
\begin{figure*}[t]
    \centering
    \begin{tikzpicture}[
        font=\sffamily\small,
        >=Latex,
        line width=0.75pt,
        box/.style={
            rounded corners=3pt,
            draw=linegray,
            fill=softgray,
            align=center,
            inner xsep=7pt,
            inner ysep=6pt
        },
        oracle/.style={box, draw=oracleblue, very thick, text width=4.20cm,
                       minimum height=3.6cm},
        explicit/.style={box, draw=black!65, text width=4.10cm,
                         minimum height=3.6cm},
        branch/.style={
            box,
            text width=4.00cm,
            minimum height=2.95cm,
            line width=0.9pt
        },
        arrow/.style={-{Latex[length=2.4mm,width=1.7mm]}, draw=linegray}
    ]
    \providecommand{\gloss}[1]{{\footnotesize\color{black!60}#1}}
    \providecommand{\sectag}[2]{{\scriptsize\bfseries\color{#1}#2}}

    \node[oracle] (oracle) {
        {\bfseries Value-oracle model}\\[1pt]
        \gloss{query \(x\), learn only \(C(x)\)}\\[5pt]
        \(\displaystyle Q_{\mathrm Q}=O(1)\)\\[2pt]
        vs.\ \(\displaystyle Q_{\mathrm C}=\Omega\!\left(\frac{n}{\log n}\right)\)\\[3pt]
        \gloss{proven quantum--classical separation}
    };

    \node[explicit, right=20mm of oracle] (explicit) {
        {\bfseries Explicit clause list}\\[1pt]
        \gloss{\(\ell\)-local clauses, kept with prob.\ \(f\)}\\[5pt]
        \(\displaystyle C=\sum_{\mu=1}^{m} C_\mu\)\\[3pt]
        \(\displaystyle m=\Theta(n d_n)=\Theta(f n^\ell)\)
    };

    \draw[arrow] (oracle) --
        node[midway, above=0.8mm, align=center,
             font=\sffamily\scriptsize\bfseries, text=black!65]
        {explicit\\realization}
        (explicit);

    \node[branch, draw=green, fill=green!7, anchor=west,
          at={($(explicit.east)+(17mm,17.8mm)$)}] (quantum) {
        \sectag{green}{Sec.~\ref{sec:cheap}}\\
        {\bfseries\color{green} Small-angle synthesis}\\[3pt]
        \gloss{rotation angle}\\
        \(\gamma=\Theta\big(1/(f n^{\ell-1})\big)\)\\[7pt]
        \(N_T=\widetilde O(n^2)\)\\
        \gloss{independent of \(\ell\)}
    };

    \node[branch, draw=classicalred, fill=classicalred!7, anchor=west,
          at={($(explicit.east)+(17mm,-17.8mm)$)}] (classical) {
        \sectag{classicalred}{Sec.~\ref{sec:leak}}\\
        {\bfseries\color{classicalred} Low-degree leakage}\\[3pt]
        \gloss{one-body coefficient}\\
        \(C=\mathrm{const}+\textstyle\sum_ih_iZ_i+\dots\)\\[7pt]
        \(\operatorname{sgn}(h_i)=(-1)^{s_i}\)\\
        \gloss{sign identifies planted solution}
    };

    \draw[arrow, draw=green]
        ([yshift=0.45pt]explicit.east)
        to[out=0, in=180, looseness=0.75] (quantum.west);
    \draw[arrow, draw=classicalred]
        ([yshift=-0.45pt]explicit.east)
        to[out=0, in=180, looseness=0.75] (classical.west);

    \end{tikzpicture}
    \caption{%
        \textbf{From value oracle to explicit clauses.}
        The quantum--classical separation of Ref.~\cite{montanaroQuantumSpeedupsSolving2025} is established in a value-oracle model, where an algorithm accesses the cost only through queries to $C(x)$ (left). For the clause-based implementation considered here, the same cost is represented explicitly as $C=\sum_\mu C_\mu$ (middle).  This representation has two consequences: the phase separator consists of many parametrically small rotations, whose non-Clifford synthesis cost reduces to $\widetilde O(n^2)$ instead of $\widetilde O(n^{\ell})$ (top right), while for the planted families of Ref.~\cite{montanaroQuantumSpeedupsSolving2025} the clause list also exposes the planted solution, through the signs of the degree-one coefficients $h_i$ (bottom right).
    }
    \label{fig:opening-oracle}
\end{figure*}

\section{Near-symmetric QAOA}
\label{sec:montanaro-summary}

A depth-\(p\) QAOA circuit~\cite{farhiQuantumApproximateOptimization2014} applies \(p\) alternating layers of a phase separator, generated by the cost function, and a mixer, generated by a transverse field. We work throughout at \(p=1\), where the resulting state is
\begin{equation}
    \ket{\psi_{\gamma,\beta}}
    =
    e^{i\beta\sum_{j=1}^{n}X_j}\;e^{i\gamma C}\;\ket{+}^{\otimes n} .
    \label{eq:qaoa-state}
\end{equation}
Here \(X_j\) is the Pauli-\(X\) operator on qubit \(j\), \(\ket{+}^{\otimes n} = 2^{-n/2} \sum_x \ket{x}\) is the uniform superposition over all \(2^n\) bit strings, and \(C\) denotes both the cost \(C:\mathbb{F}_2^n\to\mathbb{Z}\) and the diagonal operator \(C\ket{x}=C(x)\ket{x}\) it defines. We fix the mixer angle \(\beta=\pi/4\) throughout, following Ref.~\cite{montanaroQuantumSpeedupsSolving2025}, leaving \(\gamma\) as the only free parameter.

\subsection{Symmetry and phase matching}
\label{sec:phase-matching}

In the constructions of Ref.~\cite{montanaroQuantumSpeedupsSolving2025}, the cost is symmetric about a hidden solution \(s\in\mathbb{F}_2^n\). Concretely, it is invariant under a group \(G\subseteq S_n\) of permutations of the \(n\) variables (with \(S_n\) the symmetric group) acting on \(x\oplus s\),
\begin{equation}
    C\!\left(\pi(x\oplus s)\oplus s\right)=C(x),
    \qquad \pi\in G .
    \label{eq:g-symmetry}
\end{equation}
The cost is therefore constant on each orbit of $G$, so it is specified by one value per orbit. Our running example is \(G=S_n\), whose orbits are the \(n+1\) Hamming shells around \(s\), so that the cost collapses to a single function of the Hamming distance, \(C(x)=c(\abs{x\oplus s})\) with \(c:\{0,\dots,n\}\to\mathbb{Z}\). This generalizes to product symmetries such as \(S_{n_1}\times S_{n_2}\), with \(n_1+n_2=n\), which instead leads to a grid of orbits labelled by the blockwise distances. In either case, a problem over \(2^n\) bit strings is reduced to one over polynomially many orbits, which makes the depth-one analysis tractable.

For \(G=S_n\), grouping the terms of Eq.~\eqref{eq:qaoa-state} into Hamming shells gives the overlap with the planted string
\begin{equation}
    A(\gamma):=\braket{s}{\psi_{\gamma,\pi/4}}
    =
    \frac{1}{2^n}\sum_{k=0}^{n}\binom{n}{k}\,i^{k}e^{i\gamma c(k)} ,
    \label{eq:mz-overlap}
\end{equation}
and QAOA finds the planted solution with probability \(p_\star:=\abs{A(\gamma)}^2\).

The binomial factor $\binom{n}{k}$ is peaked at \(k=n/2\) with width \(O(\sqrt n)\), so the sum in Eq.~\eqref{eq:mz-overlap} depends on \(c(k)\) mainly through how the contributions from the shells near the peak at $n/2$ interfere with each other.
Defining \(\xi=k/n\), for a cost built from clauses on at most \(\ell\) variables, \(c\) has the large-\(n\) form
\begin{equation}
    c(\xi n)=n^{\ell}F(\xi)+O(n^{\ell-1}) ,
    \label{eq:profile}
\end{equation}
with profile \(F=\Theta(1)\). Scaling the angle as \(\gamma=\Gamma/n^{\ell-1}\), the summand of
Eq.~\eqref{eq:mz-overlap} carries the phase
\begin{equation}
    \frac{\pi k}{2}+\gamma c(k)
    =
    n\left[\frac{\pi}{2}\xi+\Gamma F(\xi)\right] ,
    \label{eq:total-phase}
\end{equation}
where the first term is the mixer factor \(i^k\) and the second stems from the phase separator. The dominant shells add coherently when the phase is stationary at the peak \(\xi=1/2\) of the binomial, which happens when
\begin{equation}
    \Gamma F'(1/2)=-\frac{\pi}{2} .
    \label{eq:phase-matching}
\end{equation}
We refer to this as the \emph{phase-matching condition}: the phase separator must match the phase imposed by the balanced mixer on the dominant orbits. It is solved by some \(\Gamma=\Theta(1)\) whenever \(F'(1/2)\neq0\), and for suitable profiles \(F\) this is the angle at which Ref.~\cite{montanaroQuantumSpeedupsSolving2025} proves \(\Omega(1)\) success.

We note that this generalizes beyond a single Hamming coordinate: under a product symmetry the
orbit sum runs over one normalized distance per block, each binomial peaks at \(1/2\), and
Eq.~\eqref{eq:phase-matching} becomes one stationarity condition per block, weighted by that
block's share of the variables.

In summary, phase matching fixes the product of the angle and the slope of the cost at the dominant shell. Since this slope grows as \(n^{\ell-1}\), phase matching leads to the small-angle scaling \(\gamma=\Theta(n^{1-\ell})\) that is exploited to reduce the non-Clifford cost in Sec.~\ref{sec:cheap}, while Sec.~\ref{sec:leak} leverages the fact that $F'(1/2)\neq 0$ to expose the planted solution.

\subsection{Separation in the value-oracle model}
\label{sec:quantum-classical-separation}
Ref.~\cite{montanaroQuantumSpeedupsSolving2025} shows that in the value-oracle model this mechanism generates a quantum-classical separation. For the phase-matched families above, depth-one QAOA reaches the planted solution with \(O(1)\) quantum queries to this oracle, whereas any classical algorithm needs \(\Omega(n/\log n)\) queries. The lower bound on classical query complexity rests on a counting argument. An \(S_n\)-symmetric cost takes at most \(n+1\) values, so each classical query returns \(O(\log n)\) bits, and recovering the \(n\)-bit string \(s\) requires \(\Omega(n/\log n)\) of them. This bound is known to be tight~\cite{montanaroQuantumSpeedupsSolving2025,wangDataExtractionHistogram2016, soleymani2024nonadaptive, gebhardParallelReconstructionPooled2022}.

We note that the bound holds against arbitrary classical algorithms and that local heuristics may do much worse than the general lower bound. Ref.~\cite{montanaroQuantumSpeedupsSolving2025} constructs deceptive costs where the planted string is the global minimum, but the cost decreases away from it over most of the range~\cite{farhiQuantumAdiabaticEvolution2002, muthukrishnanTunnelingSpeedupQuantum2016}. Hence, hill climbing needs exponentially many random restarts, and simulated annealing exponential time to cross a \(\Theta(n^{\ell})\) energy barrier.

\subsection{From symmetric costs to explicit CSPs by symmetry breaking}

Exactly symmetric instances are easy for classical algorithms that know the symmetry. Querying the \(n+1\) strings of Hamming weight at most one already recovers \(s\), and the algorithm behind the tight bound of Sec.~\ref{sec:quantum-classical-separation} needs only \(O(n/\log n)\) queries and polynomial time~\cite{montanaroQuantumSpeedupsSolving2025}. Ref.~\cite{montanaroQuantumSpeedupsSolving2025} therefore breaks the exact symmetry. Starting from a dense symmetric \(\ell\)-CSP with \(\Theta(n^\ell)\) local clauses, each clause is retained independently with probability \(f\). Writing \(f=d_n/n^{\ell-1}\), so that \(d_n\) is the average interaction degree, the phase-matched angle has to be rescaled as
\begin{equation}
    \gamma
    =
    \frac{\Gamma}{f n^{\ell-1}}
    =
    \frac{\Gamma}{d_n} ,
    \label{eq:mz-sparse-angle}
\end{equation}
and the constant-success guarantee survives, with constant probability over the sampled instances, as long as \(d_n=\Omega(n)\).

Ref.~\cite{montanaroQuantumSpeedupsSolving2025} benchmarks six SAT and Max-SAT solvers (including recent winners of the MaxSAT Evaluation exact tracks) on these instances. Hardness is not generic across their constructions: on the near-$S_n$-symmetric Max-$4$ and Max-$5$ families some solvers still manage to solve the instances in time linear in the number of clauses. However, strikingly, for the more complex near-$S_{n}$-symmetric and near-$S_{n/2}\times S_{n/2}$-symmetric Max-$5$ and Max-$6$ families all tested Max-SAT solvers scale exponentially, while $p=1$ QAOA retains constant success probability. This empirical evidence of an exponential speedup with $p=1$ QAOA makes these instances a natural target for an implementation on early fault-tolerant hardware.

\section{Fault-tolerant cost}
\label{sec:cheap}

We now ask what resources are required to implement these QAOA circuits fault-tolerantly. We adopt a surface-code-based fault-tolerant architecture, in which Clifford operations can be implemented using stabilizer operations such as Pauli-frame tracking and lattice surgery, while non-Clifford gates are commonly supplied through injected magic states~\cite{litinskiGameSurfaceCodes2019}. The production and consumption of these states play a central role in the space--time organization of large-scale fault-tolerant computations~\cite{litinskiGameSurfaceCodes2019,beverlandAssessingRequirementsScale2022,sandersCompilationFaultTolerantQuantum2020}. Recent protocols have substantially reduced the estimated cost of producing magic states~\cite{gidneyMagicStateCultivation2024}, but non-Clifford resources remain a useful metric for characterizing the structure and cost of fault-tolerant computation.

Arbitrary-angle rotations are particularly relevant in this respect. A generic rotation is not an element of a finite fault-tolerant gate set such as Clifford+$T$ and must instead be synthesized, with conventional unitary synthesis requiring $O(\log(1/\varepsilon))$ $T$ gates at precision $\varepsilon$~\cite{rossOptimalAncillafreeClifford+T2016}. From this perspective, the constructions of Ref.~\cite{montanaroQuantumSpeedupsSolving2025} initially appear costly. Their dense symmetric $\ell$-CSPs contain $\Theta(n^\ell)$ clauses, so a direct clause-by-clause implementation of the $p=1$ phase separator contains the same number of arbitrary-angle phase rotations. The mixer at $\beta=\pi/4$ is Clifford; moreover, even for a generic mixer angle it would contribute only $n$ single-qubit rotations. Thus, the bulk of the non-Clifford cost is concentrated in the phase-separator layer.

The crucial feature we exploit below is that these rotations have angles that shrink with the system size. Phase matching, Eq.~\eqref{eq:phase-matching}, fixes
\begin{equation}
    \gamma=\Theta(n^{1-\ell}),
\end{equation}
which changes the synthesis cost qualitatively and reduces the non-Clifford cost per circuit to $\widetilde O(n^2)$.

\subsection{Small-angle synthesis}
\label{sec:small-angle-synthesis}

Bothe \emph{et al.}~\cite{botheMoreEfficientClifford+T2026} leverage the fact that a small-angle rotation channel $\mathcal Z_\gamma(\rho):=e^{i\gamma Z}\rho\,e^{-i\gamma Z}$ is close to the identity for $\abs{\gamma}\ll1$, so that most such rotations can be replaced by the identity. They study two related mixed-synthesis schemes. In the quasi-probability formulation the target rotation channel is represented exactly as
\begin{equation}
    \mathcal Z_\gamma=\sum_i c_i\mathcal U_i,
    \qquad
    \sum_i |c_i|=1+\delta ,
    \label{eq:qp-decomposition}
\end{equation}
where the $\mathcal U_i$ are Clifford+$T$ channels and the coefficients may have either sign, so that $\delta\geq0$. In the probability formulation one instead implements
\begin{equation}
    \widetilde{\mathcal Z}_\gamma=\sum_i p_i\mathcal U_i,
    \quad
    p_i\geq0,\quad \sum_i p_i=1 ,
\end{equation}
with diamond-norm error $\|\widetilde{\mathcal Z}_\gamma-\mathcal Z_\gamma\|_\diamond=\delta$.
For equal $\delta$, the two formulations have essentially the same average $T$-count as a function of $\gamma$ and $\delta$, despite the different roles that $\delta$ plays in the two formulations.

The channels $\mathcal U_i$ and their weights depend on $\gamma$ and $\delta$. For small angles, the identity can serve as the under-rotation and carries weight $1-O(\gamma^2/\delta)$, while the remaining weight goes to Clifford+$T$ over-rotations. For $\delta\gg\gamma^2$, most rotations are therefore dropped. In the asymptotic regime
\begin{equation}
    \gamma^2\ll\delta\ll\abs{\gamma}\ll1 ,
    \label{eq:window}
\end{equation}
where the identity dominates the mixture and the over-rotation angles $O(\delta/\abs{\gamma})$ are still small, Ref.~\cite{botheMoreEfficientClifford+T2026} finds that the average $T$-count per rotation is
\begin{equation}
    N_T^{(\mathrm{rot})}(\gamma,\delta)
    =\widetilde O\!\left(\frac{\gamma^2}{\delta}\right),
    \label{eq:small-angle-cost}
\end{equation}
where the $\widetilde O$ hides the logarithmic cost of the non-identity branches. For $\delta\lesssim\gamma^2$, the identity ceases to be the optimal under-rotation and the cost crosses over to the usual angle-independent $O(\log(1/\delta))$ behavior.

\begin{figure}[t]
    \centering
    \includegraphics{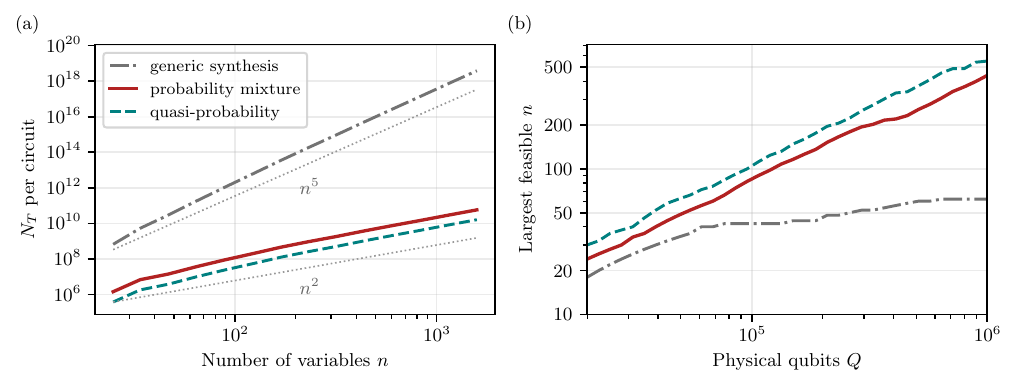}
    \caption{\textbf{Fault-tolerant cost of the depth-one QAOA }for the Max-$5$ instance of Appendix~\ref{app:instance} with success probability $p_\star$ (Eq.~\eqref{eq:max5-pstar}) and a sparsification rate $f=0.8$.
    \textbf{(a) Expected $T$-count per circuit}, $N_T$. Generic synthesis scales with the number of clauses $\Theta(n^\ell)$, whereas both small-angle conventions follow Eq.~\eqref{eq:nrot} and grow as $n^2$; the probability mixture is evaluated at $\chi$ that minimizes the total non-Clifford cost and the quasi-probability at $\chi\simeq1$.
    \textbf{(b) Largest problem size that fits a machine with $Q$ physical qubits at a physical error rate of $10^{-3}$}, within a per-circuit fault budget and a time to solution of one hour. The magic-state source and the number of factories are optimized. The quasi-probability decomposition runs at the tolerance $\chi$ that maximizes the reachable problem size $n$. Details of the resource estimate are provided in Appendix~\ref{app:resources}.%
    }
    \label{fig:ft-cost}
\end{figure}

\subsection{Non-Clifford cost per circuit}
\label{sec:rotation-count}

For the sparsified instances of Sec.~\ref{sec:montanaro-summary}, the clause count and the phase-matched angle of Eq.~\eqref{eq:mz-sparse-angle} are
\begin{equation}
    m=f\alpha n^\ell=\alpha n d_n ,
    \qquad
    \gamma=\frac{\Gamma}{d_n} ,
    \label{eq:m-def}
\end{equation}
with $d_n=fn^{\ell-1}$ the average degree and $\alpha=\Theta(1)$ a constant fixed by the construction, such that the dense parent has $\alpha n^\ell$ clauses to leading order. We count one rotation of angle $\gamma$ per clause; merging rotations (App.~\ref{app:resources}) saves only a constant factor. Additionally, let $\chi:=m\delta$ be the tolerance allocated across the layer, taken uniformly because all $m$ rotations have the same angle. We discuss the role of $\chi$ in controlling the overall error and the trade-offs between the two formulations in Sec.~\ref{sec:budget}; for now it suffices to take \(\chi=\Theta(1)\), corresponding to a fixed circuit-level error budget, so that it does not affect the asymptotic scaling with \(n\).

In the small-angle regime of Eq.~\eqref{eq:window}, Eq.~\eqref{eq:small-angle-cost} then gives the expected $T$-count per circuit
\begin{equation}
    N_T
    =m\,N_T^{(\mathrm{rot})}(\gamma,\delta)
    =m\,\widetilde O\!\left(\frac{\gamma^2}{\delta}\right)
    =\widetilde O\!\left(\frac{(m\gamma)^2}{\chi}\right)
    =\widetilde O\!\left(
        \frac{1}{\chi}
        \Big(
            \underbrace{f\alpha n^\ell}_{m}\,
            \underbrace{\tfrac{\Gamma}{f n^{\ell-1}}}_{\gamma}
        \Big)^{\!2}
      \right)
    =\widetilde O\!\left(
        \frac{\alpha^2\Gamma^2 n^2}{\chi}
      \right).
    \label{eq:nrot}
\end{equation}
The cost is therefore set by the total phase mass $m\gamma=\alpha\Gamma n$, which is linear in $n$. Hence, we find $N_T=\widetilde O(n^2)$ for any family of this form; Fig.~\ref{fig:ft-cost}(a) shows this for the running instance with $\ell=5$. The locality $\ell$ and the sparsification fraction $f$ enter only through the logarithmic branch cost hidden in $\widetilde O$, at most $O(\log(1/\delta))=O(\ell\log n)$. Depending on how the clauses are compiled, the phase separator can require non-Clifford resources beyond the rotations themselves, for instance multi-controlled gates that compute a clause indicator into an ancilla. These contribute at most an $O(\ell)$ factor per retained rotation and do not change the scaling of Eq.~\eqref{eq:nrot}. We discuss compilation details in App.~\ref{app:resources}.

It remains to check that we are in the small-angle regime of Eq.~\eqref{eq:window}. With $\delta=\chi/m$,
\begin{equation}
    \frac{\delta}{\abs{\gamma}}=\frac{\chi}{\alpha\abs{\Gamma}n} ,
    \qquad
    \frac{\delta}{\gamma^2}=\frac{\chi\,d_n}{\alpha\Gamma^2 n} ,
    \label{eq:window-check}
\end{equation}
so $\delta\ll\abs{\gamma}$ holds automatically at fixed $\chi$, while $\gamma^2\ll\delta$ requires $d_n\gg n$. At the sparsest edge of the constant-success regime of Ref.~\cite{montanaroQuantumSpeedupsSolving2025}, $d_n=\Theta(n)$, synthesis can instead sit near the crossover $\delta\sim\gamma^2$. However, at $d_n=\Theta(n)$ the total number of rotations is already $m=\Theta(n^2)$, so even the angle-independent $O(\log1/\delta)$ cost gives $N_T=\widetilde O(n^2)$. The whole regime $d_n=\Omega(n)$ in which the constant-success guarantee holds therefore has quadratic non-Clifford cost per circuit.

\subsection{Error budget and sampling overhead}
\label{sec:budget}

The role of $\chi$ differs between the two synthesis conventions. A probability mixture is approximate: the total channel error is upper bounded by $\chi$, so every measurement probability shifts by at most $O(\chi)$. Retaining a constant fraction of a target solution's ideal probability $p_\star$ therefore needs $\chi=O(p_\star)$, and since the cost falls as $\chi$ grows, the cost-minimizing choice is $\chi=\Theta(p_\star)$. With $m\gamma=\alpha\Gamma n$, Eq.~\eqref{eq:nrot} then gives $N_T=\widetilde O(n^2/p_\star)$ per circuit, and a sample complexity of $N_{\mathrm{shots}}=O(1/p_\star)$, yielding a total non-Clifford cost of
\begin{equation}
    N_T N_{\mathrm{shots}}
      =\widetilde O\!\left(
          \frac{n^2}{p_\star^2}
        \right).
    \label{eq:prob-total-cost}
\end{equation}

In contrast, the quasi-probability decomposition is exact and therefore introduces no synthesis error. For the whole layer the per-rotation norms multiply, $\Lambda=(1+\delta)^m\leq e^{\chi}$. As is customary in quasi-probability methods~\cite{temmeErrorMitigationShortDepth2017}, estimating an expectation value from a signed decomposition incurs the familiar variance overhead $\Lambda^2$. The optimization setting considered here is less demanding~\cite{barronProvableBoundsNoisefree2024}. We show in Appendix~\ref{app:unsigned} that dropping the signs while sampling branches with probability proportional to their coefficients then yields a target solution with probability
\begin{equation}
    q_\star\geq\frac{p_\star}{\Lambda},
    \label{eq:unsigned-bound}
\end{equation}
so $N_{\mathrm{shots}} = O(\Lambda/p_\star)$ shots suffice, rather than the $O(\Lambda^2)$ overhead associated with
unbiased expectation-value estimation. The total non-Clifford cost of the quasi-probability scheme is therefore
\begin{equation}
    N_T N_{\mathrm{shots}}
      =\widetilde O\!\left(
          \frac{n^2}{p_\star}\,
          \frac{e^\chi}{\chi}
        \right).
    \label{eq:qp-total-cost}
\end{equation}
For any constant $\chi$ allowed by the small-angle regime this becomes
\begin{equation}
    N_T N_{\mathrm{shots}}
      =\widetilde O\!\left(
          \frac{n^2}{p_\star}
        \right),
\end{equation}
an asymptotic factor $1/p_\star$ smaller than Eq.~\eqref{eq:prob-total-cost}. For the phase-matched families considered here $p_\star=\Theta(1)$, so this difference is only a constant factor and both totals are $\widetilde O(n^2)$. Their non-Clifford costs per circuit likewise differ only by a constant, as shown in Fig.~\ref{fig:ft-cost}(a).

The quasi-probability parameter $\chi$ is therefore tunable rather than constrained by an accuracy budget. Increasing $\chi$ reduces the non-Clifford cost per circuit as $N_T\propto \chi^{-1}$, at the price of a sampling overhead bounded by $e^\chi$. In the idealized product $e^\chi/\chi$ the minimum occurs at $\chi=1$, while choosing a larger $\chi$ can be useful when reducing the hardware requirement per circuit is more important than minimizing the total non-Clifford cost. Fig.~\ref{fig:ft-cost}(b) shows this for the running instance: on a machine of fixed size, a larger sampling overhead lets the quasi-probability decomposition fit about $30\%$ more variables than the mixture (cf. Appendix~\ref{app:resources}).

\section{Phase matching exposes the planted solution}
\label{sec:leak}

Both the satisfiability solvers tested in Ref.~\cite{montanaroQuantumSpeedupsSolving2025} and the resource estimates of Sec.~\ref{sec:cheap} require a stronger access model than the value oracle used in Sec.~\ref{sec:montanaro-summary}. We implement the phase separator through an explicit local decomposition
\[
C(x)=\sum_{\mu=1}^{m} C_\mu(x),
\]
since each clause must be compiled into a corresponding rotation. We next ask what use classical algorithms can make of the clause list required for compilation.

\subsection{Low-degree recovery of the planted solution}
\label{sec:low-degree-recovery}

Consider first the fully symmetric parent
\[
    C(x)=c(\abs{x\oplus s}),
    \qquad
    c(k)=n^\ell F(k/n)+O(n^{\ell-1}).
\]
In the Ising expansion
\[
    C=\mathrm{const}+\sum_i h_i Z_i+\cdots ,
\]
the one-body coefficient $h_i$ measures the average change in cost obtained by flipping
bit $i$. Concretely,
\begin{equation}
    h_i
    =
    \frac{(-1)^{s_i}}{2}
    \left(
        \mathbb E[C\mid x_i=s_i]
        -
        \mathbb E[C\mid x_i\neq s_i]
    \right).
    \label{eq:field-def}
\end{equation}
Conditioned on either value of $x_i$, the Hamming distance to the planted solution contributed by the remaining
$n-1$ bits is distributed as $K\sim\mathrm{Bin}(n-1,1/2)$. Hence
\begin{equation}
    h_i
    =
    \frac{(-1)^{s_i}}{2}\,
    \mathbb E\!\left[c(K)-c(K+1)\right].
\end{equation}
For the fixed-locality symmetric CSPs considered here, the cost changes from one shell to the
next by
\[
    c(k+1)-c(k)
    =
    n^{\ell-1}F'(k/n)+O(n^{\ell-2}).
\]
A random bit string lies near the peak of the binomial, which is symmetric about $K\approx n/2$, so
\begin{equation}
    h_i
    =
    -\frac{(-1)^{s_i}}{2}\,
    n^{\ell-1}F'(1/2)
    +O(n^{\ell-2}).
    \label{eq:field-slope}
\end{equation}
Thus, all one-body coefficients have the same leading magnitude and differ only in sign through the planted bit $s_i$. Whenever $F'(1/2)\neq0$, their signs determine the planted solution.

The same derivative condition appears in the phase-matching condition. Indeed, since
$$
    \gamma n^{\ell-1}F'(1/2)=-\frac{\pi}{2},
    \qquad
    \gamma=\frac{\Gamma}{n^{\ell-1}},
$$
Eq.~\eqref{eq:field-slope} implies
\begin{equation}
    \gamma h_i
    =
    \frac{\pi}{4}(-1)^{s_i}+O(n^{-1}).
    \label{eq:field-pinning}
\end{equation}
Therefore, the same slope of the cost profile that is required for phase matching fixes a macroscopic degree-one signal that directly reveals the planted solution.

Crucially, this signal is easily accessible from the clause list. If
$C=\sum_{\mu=1}^m C_\mu$, then
\[
    h_i=\sum_{\mu=1}^m a_{\mu i},
\]
where $a_{\mu i}$ is the coefficient of $Z_i$ contributed by clause $\mu$, which vanishes unless $i$ appears in that clause. Hence $h_i$ can be accumulated in a single pass over the clause list, making the recovery of the planted solution straightforward.

We note that the argument carries over to the product symmetries of Sec.~\ref{sec:phase-matching}, which include the near-$S_{n/2}\times S_{n/2}$ instances for which the runtime of solvers tested in Ref.~\cite{montanaroQuantumSpeedupsSolving2025} scales exponentially. For $G=S_{n_1}\times S_{n_2}$, write $\xi_a=k_a/n_a$ for the normalized distance to $s$ in block $a$, so that
$c=n^\ell F(\xi_1,\xi_2)+O(n^{\ell-1})$. Flipping a bit $i$ in block $a$ changes only $k_a$,
and Eq.~\eqref{eq:field-slope} becomes
\begin{equation}
    h_i
    =
    -\frac{(-1)^{s_i}}{2}\,
    \frac{n^{\ell}}{n_a}\,
    \partial_aF(\tfrac12,\tfrac12)
    +O(n^{\ell-2}).
\end{equation}
Phase matching now holds block by block, $\Gamma\,\partial_aF(\tfrac12,\tfrac12)=-\tfrac{\pi}{2}n_a/n$ (cf. App.~\ref{app:instance}), so Eq.~\eqref{eq:field-pinning} holds for every $i$ with the same constant. The signs of $h_i$ therefore reveal $s$ without knowledge of the block partition.

\subsection{Recovery after sparsification}
\label{sec:sparse-recovery}
We now show that this signal persists under random sparsification. As before, we retain each clause independently with probability $f$ and write
\begin{equation}
    \widetilde h_i
    =
    \sum_\mu w_\mu a_{\mu i},
    \qquad
    w_\mu\sim\mathrm{Bernoulli}(f),
\end{equation}
with the expected value $\mathbb E[\widetilde{h}_i]=f h_i$. With the average degree $d_n=fn^{\ell-1}$ of Eq.~\eqref{eq:m-def}, Eq.~\eqref{eq:field-slope} shows that the mean signed signal has magnitude
${
    |\mathbb E[\widetilde h_i]|
    =
    \Theta(d_n)
}$. The sign of $\widetilde h_i$ misreports $s_i$ only if a fluctuation flips it, which requires a fluctuation as large as the mean, $\Theta(d_n)$.

Each of the $\Theta(n^{\ell-1})$ clauses of the dense parent that contain $i$ contributes
$a_{\mu i}=\pm2^{-\ell}$. Hence
\begin{equation}
    \operatorname{Var}(\widetilde{h}_i)
    =
    f(1-f)\sum_\mu a_{\mu i}^2 \\
    =
    O(f n^{\ell-1})
    =
    O(d_n),
\end{equation}
so a typical fluctuation is only $O(\sqrt{d_n})$, far below the $\Theta(d_n)$ needed to flip the sign. With the estimator
\[
    (-1)^{\widehat s_i}
    :=
    \operatorname{sgn}(\gamma\widetilde{h}_i),
\]
Bernstein's inequality gives, for some family-dependent constant $b>0$,
\begin{equation}
    \Pr(\widehat s_i\neq s_i)
    \le
    e^{-b d_n},
    \label{eq:single-bit-recovery}
\end{equation}
and invoking a union bound over all $n$ variables yields
\begin{equation}
    \Pr(\widehat{\bm s}\neq\bm s)
    \le
    n e^{-b d_n}.
    \label{eq:recovery}
\end{equation}

Consequently, any $d_n$ that is a sufficiently large multiple of $\log n$ already makes the recovery probability tend to one. By contrast, the constant-success QAOA guarantee of Ref.~\cite{montanaroQuantumSpeedupsSolving2025} requires $d_n=\Omega(n)$. The instances covered by that guarantee therefore lie well inside the regime in which the planted string can be recovered from the explicit degree-one structure.

\section{Deceptive near-symmetric instances without a planted solution}
\label{sec:escape}

In Sec.~\ref{sec:leak}, the local fields of the explicit instances revealed the planted solution. We now drop the planted solution and construct instances whose exact optimization is NP-hard, but on which depth-one QAOA still outperforms structure-unaware classical algorithms at the same small-angle synthesis cost as before.

\subsection{Construction}

To this end, we replace the planted string by an exponentially large set of optimal strings and plant an NP-hard problem in that set. The cost function takes the form $C_\eta=C_0+\eta W$ with $C_0$ the (near-)$S_n$-symmetric part  that selects an exponentially large (near-)degenerate optimal shell. The frustrated term $\eta W$ is chosen such that it keeps every global minimum on the optimal shell and makes finding the optimum within the shell NP-hard.

Let us define the normalized Hamming weight $\xi=k/n$ for $k=\abs{x}$. We take $C_0$ to be a weighted sum of $\ell$-clauses that depends only on $k$, $C_0(x)=c(k)$, with profile $F$ as in Eq.~\eqref{eq:profile}, and require that $F$ has its global minimum at an interior point $0<\xi_{\rm opt}<1$, and that $F'(\tfrac12)\neq 0$. The minima of $C_0$ then form the optimal shell $k_{\rm opt}\approx\xi_{\rm opt}n$, which is exponentially degenerate, with $\binom{n}{k_{\rm opt}}=e^{\Theta(n)}$ strings.

On the optimal shell we plant an NP-hard problem by adding a frustrated two-body term
\[
    W=\sum_{\{i,j\}\in E}J_{ij}Z_iZ_j,
    \qquad
    J_{ij}=\pm1 .
\]
Let $\Delta_n$ be the gap between the optimal shell and the nearest suboptimal shell. Since $c(k)$ is integer valued $\Delta_n\geq1$ and $\Delta_n=O(n^{\ell-2})$. If
$2\eta\norm{W} < \Delta_n$ e.g. when $\eta = 1/(2\abs{E}+1)$,
 every minimum of $C_\eta$ lies on the optimal shell.
Therefore, minimizing $C_\eta$ is an Ising ground-state problem at fixed Hamming weight. This problem is NP-hard by a reduction from maximum cut~\cite{gareySimplifiedNPcompleteGraph1976}.

\subsection{Depth-one QAOA}
\label{sec:escape-qaoa}

We now analyze how depth-one QAOA finds the optimal shell $\xi_{\mathrm{opt}}$. In the initial state $\ket{+}^{\otimes n}$, the Hamming weight $k$ is distributed as $k_{\rm initial}\sim\mathrm{Bin}(n,\tfrac12)$, with mean $n/2$ and variance $n/4$, so the state is concentrated on the shells with $\abs{k-\tfrac n2}=O(\sqrt n)$. Expanding $c(k)$ about $k=n/2$ with Eq.~\eqref{eq:profile} gives, on these shells,
\begin{equation}
    \gamma c(k)=\gamma c(\tfrac n2)+\varphi\,(k-\tfrac n2)+\frac{\kappa}{n}\,(k-\tfrac n2)^2+O(n^{-1/2}),
    \qquad
    \varphi:=\gamma n^{\ell-1}F'(\tfrac12),
    \quad
    \kappa:=\tfrac12\gamma n^{\ell-1}F''(\tfrac12).
    \label{eq:gamma-c-expansion}
\end{equation}
At $\gamma=\Theta(n^{1-\ell})$, both $\varphi$ and $\kappa$ are of order unity. Since $k=|x|$, the linear term $\varphi(k-\tfrac n2)$ is a sum of single-bit phases. On its own it creates a product state in which, up to a global phase, every qubit is in $(\ket{0}_j+e^{i\varphi}\ket{1}_j)/\sqrt2$.
The mixer layer with $\beta=\pi/4$ also acts on each qubit separately: it keeps a bit with amplitude $1/\sqrt2$ and flips it with amplitude $i/\sqrt2$, as in Eq.~\eqref{eq:mz-overlap}. Measuring a qubit thus gives outcome $1$ with probability
\[
    \frac{\abs{e^{i\varphi}+i}^2}{4}=\tfrac12(1+\sin\varphi).
\]

The quadratic term in Eq.~\eqref{eq:gamma-c-expansion} instead couples all pairs of bits. It makes the phase per bit depend on the shell, $\varphi\to\varphi+2\kappa(k-\tfrac n2)/n$, so parts of the initial state on different shells acquire slightly different phases per bit. These shifts are of order $n^{-1/2}$ and average to zero over $k_{\rm initial}$, so to leading order they do not move the mean of the output shell,
\begin{equation}
    \frac{\langle k_{\rm out}\rangle}{n}=\tfrac12(1+\sin\varphi).
    \label{eq:kout-mean}
\end{equation}
These shifts correlate each pair of bits only at order $1/n$, so $\operatorname{Var}(k_{\rm out})=O(n)$ and $k_{\rm out}$ concentrates around this mean, with fluctuations of order $\sqrt n$.

Choosing the angle such that the output is centered on the optimal shell, $\langle k_{\rm out}\rangle=k_{\rm opt}$, gives $\sin\varphi=2\xi_{\rm opt}-1$ to leading order, independently of $\kappa$, yielding $\gamma=\Theta(n^{1-\ell})$. Therefore, the fault-tolerant compilation with a non-Clifford cost of $\widetilde O(n^2)$ Sec.~\ref{sec:cheap} also applies here. Further, for $\xi_{\rm opt}=0$ this reduces to the phase-matching condition of Eq.~\eqref{eq:phase-matching}. Finally, the frustrated term perturbs the depth-one state by at most $\abs{\gamma}\eta\norm{W}=O(\eta n^{3-\ell}),$ so it is asymptotically invisible to depth-one QAOA. Consequently, QAOA finds the optimal shell to leading order, but not the exact optimum within it.

The QAOA mechanism is also robust to sparsification. Keeping each clause independently with probability $f=\Theta(1)$ leaves the average cost $fc(k)$, which is compensated by the rescaled angle $\gamma/f$. Similarly to Ref.~\cite{montanaroQuantumSpeedupsSolving2025}, the remaining random part of the phase changes the depth-one state of a typical instance by only $O(n^{1-\ell/2})$ in norm, so the output shell is unchanged to leading order.

\subsection{Comparison to classical solvers}
\label{sec:classical-solvers-unplanted}

\begin{figure}[t]
    \centering
    \includegraphics{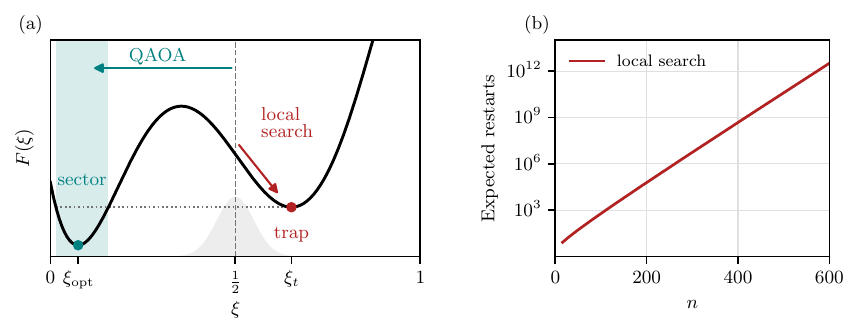}
    \caption{\textbf{Near-symmetric instances without planting.} \textbf{(a) Cost function profile} $F(\xi)$ of the cost function used in Sec.~\ref{sec:classical-solvers-unplanted}. The shaded band is the sector: the shells on the optimum's side of the barrier (the local maximum) that are cheaper than the trap at $\xi_t$ (dotted line). A random string lies near $\xi=\tfrac12$ (grey: distribution of $\xi$ for a uniformly random string at $n=100$, arbitrary scale), on the trap's side of the barrier, so local search from it ends in the trap with high probability. Depth-one QAOA at the optimized angle centers its output on the optimal shell $\xi_{\rm opt}$. \textbf{(b) Expected number of restarts of local search} from uniformly random initial strings until one ends in the sector.}
    \label{fig:single-profile}
\end{figure}

As for the planted instances, the degree-one attack of Sec.~\ref{sec:low-degree-recovery} also applies here: the local fields are all equal, $h_i\propto F'(\tfrac12)$, so the degree-one part of the cost depends only on the Hamming weight, and a scan over the $n+1$ weights locates the optimal shell. What remains hard, also for QAOA, is the NP-hard problem within the shell. We therefore compare depth-one QAOA with general-purpose algorithms that do not use this structure, local search and MaxSAT solvers, in their ability to find the optimal shell.

For this comparison we choose $C_0$ such that its landscape is deceptive. We take $\ell=5$ and give each clause the weight $w_a$, where $a$ is the number of ones in the assignment that violates it. With all $5$-subsets present, the cost and its profile are
\[
    c(k)=\sum_{a=0}^{5}w_a\binom{k}{a}\binom{n-k}{5-a},
    \qquad
    F(\xi)=\sum_{a=0}^{5}\frac{w_a}{a!\,(5-a)!}\,\xi^a(1-\xi)^{5-a}.
\]
For $w=(w_0,\dots,w_5)=(7,4,15,0,9,9)$, $F$ has its global minimum at $\xi_{\rm opt}\approx0.08$ (see Fig.~\ref{fig:single-profile}(a)), and another local minimum, the trap, at $\xi_t>\tfrac12$, separated from the global optimum by a barrier. The sector marked in Fig.~\ref{fig:single-profile}(a) comprises the shells on the optimum's side of the barrier that are cheaper than the trap. A random string has $\xi\approx\tfrac12$ (gray in Fig.~\ref{fig:single-profile}(a)), so single-bit descent increases $\xi$ and terminates in the trap.

Classical algorithms that do not use the structure remain trapped behind the barrier with high probability. Local descent requires $e^{\Theta(n)}$ restarts (cf. Fig.~\ref{fig:single-profile}(b)), since a random start lies on the optimum's side of the barrier only with exponentially vanishing probability, and leaving the trap requires climbing a barrier of order $n^\ell$ over $\Theta(n)$ bit flips.

General-purpose MaxSAT solvers get trapped already at moderate sizes. Aperture~\cite{nadel2026aperture}, which won the weighted anytime track of the MaxSAT Evaluation 2026~\cite{maxsat2026} and combines local search with SAT-based optimization, reaches the sector within 30 minutes on every instance up to $n=28$ but fails on most at $n\geq 40$. Failed runs reach the trap shell within seconds and stay there, even when extending the budget to 90 minutes. Without the trap, the solver finds the optimal shell of instances of the same size within seconds, so the failures are caused by the trap, rather than by the instance size. SPB-MaxSAT-c~\cite{jin2024combining}, the 2024 winner of the same track~\cite{maxsat2024}, is instead built around local search with dynamic clause weighting; it aborts on memory ($\geq 16$~GB) within a minute on most instances.

As shown in Sec.~\ref{sec:escape-qaoa}, depth-one QAOA instead centres its output on the optimal shell at a small angle, $\gamma=\Theta(n^{1-\ell})$, and hence at the $\widetilde O(n^2)$ non-Clifford cost per circuit of Sec.~\ref{sec:cheap}. Thus, the empirical speedup over local descent and general-purpose solvers reported for the planted instances~\cite{montanaroQuantumSpeedupsSolving2025} carries over to our unplanted setting.

\section{Discussion and outlook}
\label{sec:discussion-outlook}

The starting point for this work is a compelling empirical observation about a shallow quantum algorithm: Montanaro and Zhou identified explicit near-symmetric constraint-satisfaction problems for which depth-one QAOA finds the planted solution with constant probability, while the strongest classical solvers tested on their hardest families exhibit apparently exponential runtime scaling~\cite{montanaroQuantumSpeedupsSolving2025}. Although this does not establish a rigorous exponential quantum advantage, it makes these instances a particularly interesting test case: if an empirical low-depth speedup of this kind survives increasingly strong classical attacks, can the corresponding quantum computation be implemented at a realistic fault-tolerant cost?

The phase-matching mechanism that gives depth-one QAOA a constant success probability forces the phase-separator angle to scale as $\gamma=\Theta(n^{1-\ell})$. We showed that this dramatically reduces the non-Clifford cost of the circuits. The $\Theta(n^\ell)$ clauses are offset by the phase separator angle of order $n^{1-\ell}$, such that small-angle synthesis lowers the non-Clifford cost per circuit to $\widetilde O(n^2)$, independent of the locality $\ell$. As Fig.~\ref{fig:ft-cost} shows, this substantially lowers the resources required for a practical fault-tolerant implementation of these circuits. This places these shallow, small-angle circuits in a markedly different resource regime from recent end-to-end estimates for QAOA on less structured optimization problems; for example, Ref.~\cite{omanakuttanThresholdFaulttolerantQuantum2025} finds a crossover for random 8-SAT using \(p=623\) QAOA with amplitude amplification at tens of millions of physical qubits under similar hardware assumptions.

We show that the small-angle regime also enables a trade-off between non-Clifford circuit cost and sampling overhead. Because the quasi-probability decomposition is exact, circuit cost can be reduced without introducing synthesis error, at the expense of additional samples. For classically verifiable solutions the signs can simply be discarded at a quadratically smaller sampling overhead than in expectation-value estimation. This yields an asymptotic factor $1/p_\star$ improvement over the probability mixture, which is constant here but may become significant when the success probability decreases with $n$, as is the case for many instances of practical interest~\cite{boulebnaneSolvingBooleanSatisfiability2024}.

For the planted families of Ref.~\cite{montanaroQuantumSpeedupsSolving2025}, we also show that phase matching fixes the signs of the degree-one Fourier coefficients so that they expose the planted string, so a linear-time classical attack succeeds wherever constant QAOA success is guaranteed. The small angle, however, does not require a planted solution. We construct unplanted NP-hard problem families in which the small angle and the $\widetilde O(n^2)$ cost persist for shallow QAOA implementations, while local search and general-purpose MaxSAT solvers increasingly fail to reach the optimal sector as $n$ grows.

The fact that phase matching yields both the small angle and the classical exposure raises the question of when small-angle synthesis is useful in optimization beyond these constructions. The small-angle regime of Eq.~\eqref{eq:window} requires $m\gamma^2\ll\chi$, and at depth one a useful angle gives a single bit flip a phase change of order one. If the clauses on a variable add coherently, this change is of order the degree $d_n$ and $m\gamma^2=\Theta(n/d_n)$, as for the instances studied here. If they add with random signs, it is of order $\sqrt{d_n}$ and $m\gamma^2=\Theta(n)$. The optimal angles of standard benchmarks follow the second scaling: they approach size-independent values for fixed-degree regular MaxCut~\cite{wurtzFixedangleConjecturesQuantum2021} and scale as $\Theta(d_n^{-1/2})$ in the large-degree, locally tree-like limit~\cite{wyboMissingPuzzlePieces2025}. Small angles at depth one therefore require coherent local fields, whose mean over a random string is the degree-one Fourier coefficient. In both the planted and the unplanted families, the property that makes the circuit cheap is thus the one that exposes the solution or the sector to classical methods.

However, this link need not hold at larger depth. With several layers, as in digitized evolutions with many short steps, the phase of a bit flip can build up over layers, so small angles no longer require coherent local fields. Whether one can exploit this freedom to obtain circuits that are simultaneously fault-tolerantly cheap, resistant to low-degree classical attacks, and sensitive to genuinely hard microscopic structure is, in our view, the central open question raised by these results.

\section*{Acknowledgments}
We thank Miha Papi\v c for discussions on sampling using the quasi-probability decomposition. We used Claude Opus 5.5 (Anthropic) for assistance with analytical derivations, numerical experiments and the editing of this manuscript. All AI-assisted outputs were reviewed and verified by the authors, who take full responsibility for the scientific content of this work.

\appendix

\section{The Max-5 running instance}
\label{app:instance}

The running example for the numbers quoted in the main text Fig.~\ref{fig:ft-cost} is the near-\(S_{n/2}\times S_{n/2}\)-symmetric Max-\(5\) instance of Ref.~\cite{montanaroQuantumSpeedupsSolving2025}. For this instance all tested Max-SAT solvers exhibit apparently exponential runtime scaling~\cite{montanaroQuantumSpeedupsSolving2025}. Its locality is \(\ell=5\), and the \(n\) variables are split into two blocks of \(n/2\). Counting the clauses of this instance with the multiplicities of Ref.~\cite{montanaroQuantumSpeedupsSolving2025}, the total number is \(m=f\alpha n^{\ell}\) to leading order, with \(\alpha=3/32\). We use $f=0.8$ throughout.

Writing \(\xi_a=k_a/n_a\) for the normalized Hamming distance within each block, the cost takes
the form of Eq.~\eqref{eq:profile} with profile
\begin{equation}
    F(\xi_1,\xi_2)
    =
    \frac{\xi_1(1-\xi_1)^4+\xi_2(1-\xi_2)^4}{32}
    +
    \frac{(\xi_1+\xi_2)(1-\xi_1)^2(1-\xi_2)^2}{64} .
    \label{eq:max5-profile}
\end{equation}
Phase matching, Eq.~\eqref{eq:phase-matching}, becomes one stationarity condition per block, weighted by its share \(\tfrac12\) of the variables, \(\tfrac{\pi}{4}+\Gamma\,\partial_aF(\xi_{\mathrm{c}})=0\) at \(\xi_{\mathrm{c}}=(1/2,1/2)\); the two coincide because the construction is symmetrized over the blocks. With \(\partial_aF(\xi_{\mathrm{c}})=-9/1024\) this gives
\begin{equation}
    \Gamma=\frac{256\pi}{9} ,
\end{equation}
so \(\alpha\Gamma=8\pi/3\) and Eq.~\eqref{eq:nrot} reads
\(N_T=\widetilde O\big(\tfrac{64\pi^2}{9}\,n^2/\chi\big)\), plotted in Fig.~\ref{fig:ft-cost}.

Evaluating the orbit sum at \(\xi_{\mathrm{c}}\) by the saddle-point method gives
\begin{equation}
    p_\star=\frac{4}{\abs{\det\mathcal H}} ,
    \qquad
    \mathcal H_{ab}=-2\delta_{ij}+i\,\Gamma\,\partial_a\partial_b F(\xi_{\mathrm{c}}) ,
    \label{eq:saddle-pstar}
\end{equation}
with $\mathcal{H}$ the Hessian of the exponent at \(\xi_{\mathrm{c}}\). For the profile of Eq.~\eqref{eq:max5-profile} we find
\begin{equation}
    p_\star=\frac{27}{\sqrt{(27-\pi^2)^2+144\pi^2}}\approx 0.652 .
    \label{eq:max5-pstar}
\end{equation}
We use this asymptotic value throughout.

\section{Unsigned sampling from a quasi-probability decomposition}
\label{app:unsigned}

Here we explain how a signed decomposition can be sampled without tracking its signs when the goal is to find a solution rather than to estimate an expectation value, as used in Sec.~\ref{sec:budget}.

Following Ref.~\cite{botheMoreEfficientClifford+T2026}, each of the $m$ rotations in the layer is replaced by the exact mixture of Eq.~\eqref{eq:qp-decomposition}, with tolerance $\delta=\chi/m$ as in Sec.~\ref{sec:rotation-count}. Drawing one term per rotation independently yields a Clifford+$T$ circuit; index these joint draws by $j$, and let $\mathcal U_j$ be the resulting circuit and $c_j$ the product of the coefficients drawn. The ideal channel is then $\sum_j c_j\mathcal U_j$, with
\begin{equation}
    \Lambda:=\sum_j |c_j|=(1+\delta)^m\leq e^{m\delta}=e^{\chi} .
\end{equation}

Estimating an expectation value from such a decomposition requires weighting each sample by $\operatorname{sign}(c_j)\Lambda$, which is what inflates the variance by $\Lambda^2$. Here no estimator is formed as each sampled bit string is checked classically and either accepted as a solution or discarded. The only question is how often the unsigned mixture returns one. Write $p_\star$ for the probability that the exact circuit does so, and let
\begin{equation}
    r_j=P(x_\star\mid\mathcal U_j)
\end{equation}
be the probability of obtaining a target solution $x_\star$ from branch $j$. Sampling branches according to
$|c_j|/\Lambda$ while discarding their signs produces the target with probability
\begin{equation}
    q_\star=\frac{1}{\Lambda}\sum_j |c_j|r_j .
\end{equation}
Exactness of the signed decomposition and the triangle inequality imply
\begin{equation}
    p_\star
      =|\sum_j c_j r_j|
      \leq\sum_j |c_j|r_j
      =\Lambda q_\star ,
\end{equation}
and hence Eq.~\eqref{eq:unsigned-bound}. Thus $O(\Lambda/p_\star)$ samples suffice to encounter a
target solution, rather than the $O(\Lambda^2)$ overhead associated with unbiased
expectation-value estimation. This distinction between estimating expectation values and
extracting good samples also appears in the noise-mitigation setting of
Ref.~\cite{barronProvableBoundsNoisefree2024}. It holds for any optimization problem with
classically verifiable solutions, independently of the particular construction studied here.

\section{Compilation and resource estimate}
\label{app:resources}

Here we compile the phase separator of the example instance of Appendix~\ref{app:instance} and estimate its fault-tolerant resources. A clause of order $\ell$ can be compiled either by expanding it into its $2^\ell-1$ Pauli-$Z$ strings, whose CNOT ladders are Clifford, or by computing its indicator into an ancilla with $\ell-1$ temporary-AND gadgets of $4$ $T$ gates each~\cite{gidneyHalvingCostQuantum2018}. The expansion changes the total phase mass $m\gamma$ only by a factor $1-2^{-\ell}$. In the gadget route a clause whose small-angle rotation is replaced by the identity is dropped together with its gadgets, so the extra cost is $4(\ell-1)$ $T$ gates per retained rotation. We use the expansion, and collect all contributions to the same Pauli support into one rotation. Their signs partly cancel, which lowers the $T$-count by a constant factor without changing the headline scaling of Eq.~\eqref{eq:nrot}. Fig.~\ref{fig:ft-cost}(b) includes this grouping, while Fig.~\ref{fig:ft-cost}(a) does not.

We translate $N_T$ into surface-code resources assuming a physical error rate of $10^{-3}$, a \SI{1}{\micro\second} cycle and a \SI{10}{\micro\second} reaction time. As magic-state sources we consider either cultivation~\cite{gidneyMagicStateCultivation2024}, with $1.5\times10^3$ physical qubits per unit and error rate $p_m=2\times10^{-9}$, or the distillation factories of Ref.~\cite{litinskiMagicStateDistillation2019}, with between $8\times10^3$ and $7\times10^4$ physical qubits and $p_m$ between $6\times10^{-10}$ and $5\times10^{-20}$. The runtime is set by how long $F$ factories need to produce the $N_T$ magic states, and cannot be shorter than one reaction time per layer of four $T$ gates. The data qubits idle for the whole run, including while they wait for magic states, and the code distance follows from this space-time volume~\cite{fowlerLowOverheadQuantum2019}. Each of the $n$ logical qubits occupies $2d^2$ physical qubits, with a (conservative) factor $1.5$ for routing. We choose the magic-state source and the number of factories $F$ to minimize the total physical-qubit count, as we are targeting the qubit-constrained early fault-tolerant regime.

Fig.~\ref{fig:ft-cost}(b) shows the largest problem size that fits a machine of $Q$ physical qubits. Each circuit must meet the fault budget $\varepsilon_m+\varepsilon_L\leq0.1$, where $\varepsilon_m=N_Tp_m$ is the total magic-state error and $\varepsilon_L$ the logical memory error. Additionally, we require the time to solution, defined as the circuit time multiplied by the expected number of samples, to remain below one hour. The synthesis tolerance does not enter the fault budget: since the outputs are checked classically, it appears only as a sampling overhead.

Small-angle synthesis is limited mainly by space: the data patches, about $3d^2n$ physical qubits, fill the machine. Above a few $10^5$ qubits, the one-hour cap also reduces the reach.
Generic synthesis, whose $T$-count grows as $N_T\propto n^5$, is limited by its $T$-count instead. With cultivation, the magic-state error $N_Tp_m$ exceeds the fault budget beyond $n\approx42$ on any machine, since no code distance reduces it. Distillation lowers $p_m$, but the reach then stops at $n\approx62$, where one reaction time per layer of four $T$ gates already fills the one-hour cap.

Cultivation remains the best source for the quasi-probability decomposition throughout, whereas generic synthesis switches to distillation near $1.5\times10^5$ qubits and the probability mixture beyond a few $10^5$. The probability mixture and generic synthesis are best run at a synthesis error of about $p_\star/2$, which halves the success probability and hence doubles the number of samples; tuning this choice further gains at most $10\%$. The quasi-probability decomposition instead profits from a large tolerance: with sampling overheads ranging from $10$ to $10^4$ depending on $n$, its circuits become short enough to lower the code distance, while all samples still fit in an hour.

\bibliography{refs_zotero,refs_misc}

\end{document}